# Nodal Frequency Performance of Power Networks

Huisheng Gao, Hui Yuan
Huanhai Xin[*], Linbin Huang
College of Electrical Engineering
Zhejiang University
Hangzhou, China
[*]xinhh@zju.edu.cn

Changyou Feng
Dispatching and Control Center
National Electric Power
Beijing, China

*Abstract*—This paper investigates how a disturbance in the power network affects the nodal frequencies of certain network buses. To begin with, we show that the inertia of a single generator is in inverse proportion to the initial rate of change of frequency (RoCoF) under disturbances. Then, we present how the initial RoCoF of the nodal frequencies are related to the inertia constants of multiple generators in a power network, which leads to a performance metric to analyze nodal frequency performance. To be specific, the proposed metric evaluates the impact of disturbances on the nodal frequency performance. The validity and effectiveness of the proposed metric are illustrated via simulations on a multi-machine power system.

*Index Terms*- Inertia, nodal frequency performance, multi-machine power system.

## I. INTRODUCTION

With the increasing integration of renewable energy in modern power systems, the synthetical inertia of the system decreases dramatically mainly because the typical control of renewable generators, e.g., phase-locked loop (PLL)-based control, shows zero equivalent inertia. This is challenging the frequency stability of modern power systems [1]-[3], and therefore the assessment of the power system's frequency performance has attracted great attentions.

To evaluate the system's frequency performance, some metrics have been proposed in [4]-[6]. Reference [4]-[5] used system $\mathcal{H}_2$ norms which accounts for network coherency to optimize the inertia allocation. Reference [6] used two kinds of norms to evaluate the frequency performance, which quantify the system frequency by the $\mathcal{L}_\infty$ norm and quantify inter-area oscillations by $\mathcal{L}_2$ norm. These proposed metrics in [4]-[6] mainly focused on the system's global frequency performance. However, the nodal frequency performance of a certain bus inside the network, has rarely been analyzed.

In fact, even if the global frequency performance satisfies the grid requirement, the nodal frequency performance of a certain bus may still be poor. For example, consider a small-capacity diesel generator which is connected to a high-inertia ac grid through long transmission lines. In this case, although the global frequency performance of the grid is strong enough, the diesel generator's nodal frequency may fluctuate seriously due to long transmission lines.

To fill this gap, this paper proposes a metric to evaluate the nodal frequency performance for arbitrary network bus in a multi-machine power system. Firstly, we analyze the relation between the inertia and the initial RoCoF of a generator. Then, in analogy with this relation, the metric is proposed to evaluate the nodal frequency performance for arbitrary network bus. Simulation results verify the validity of the proposed metric on a multi-machine power system.

## II. RELATION BETWEEN INITIAL ROCOF AND INERITA OF A GENERATOR

Considering a generator's swing equation [7]

$$\begin{cases} \dfrac{d\delta}{dt} = \omega_0 \omega \\ J\dfrac{d\omega}{dt} = P_T - P_E - D(\omega-1) + u \end{cases} \quad (1)$$

where $\omega$ and $\delta$ are the rotor speed and power angle, $\omega_0$ is the rated angular frequency, $J$ is the inertia, $D$ is the damping coefficient, $u$ is the power disturbance, $P_T$ and $P_E$ are the mechanical power and active power of the generator. $P_E$ can be obtained by power flow equation. $P_T$ is given as

$$T\dfrac{dP_T}{dt} = -K(\omega-1) - P_T \quad (2)$$

where $T$ is turbine time constant and $K$ is droop coefficient.

Consider that a power step disturbance $u = P_1$ occurs when the generator operates in the steady state, in which $P_1$ is the amplitude of disturbance. Since $P_T$, $\omega$ and $\delta$ are state variables, and $P_E$ is a function of $\delta$, these variables will not change at $t = 0^+$. So, it can be obtained by (1) that the initial RoCoF at $t = 0^+$ is given as

$$\left.\dfrac{d\omega}{dt}\right|_{t=0^+} = \dfrac{P_1}{J} \quad (3)$$

It is noteworthy that equation (3) shows that the initial RoCoF of the generator's internal bus is in inverse proportion to the generator's inertia, which is independent of the rest of power system's dynamics in a multi-machine power system.

This work is supported by 2019 Science and Technology Project of State Grid Power Company Limited ("Study on the critical indices and Real-time Evaluation Methods for Large-scale Power System Operation", SGJS0000DKJS1900265).

## III. Proposed Metric for Nodal Frequency Performance Evaluation in a Power Network

In this section, we first deduce the initial RoCoF for network buses under a power step disturbance. Then, in analogy with the relation between inertia and the initial RoCoF of the generator, a metric is proposed to evaluate the nodal frequency performance for the network buses. Finally, a two-generators system, as an example, is conducted to introduce the metric.

### A. Modeling and Derivation

Considering a multi-machine network with $n+m+a$ buses: the first $n$ buses are the generators' internal buses, the buses $n+1,\ldots,n+m$ are load buses, the buses $n+m+1,\ldots,n+m+a$ are passive buses. Note that the load buses and passive buses are both the network buses. All loads are assumed as constant power loads. The power flow equations are given as

$$\begin{cases} P_{Ei} = E_i \sum_{j \in i} B_{ij} V_j \sin(\delta_i - \theta_j), & i=1,\ldots,n \\ -P_{Li} = V_i \sum_{j \in i} B_{ij} E_j \sin(\theta_i - \delta_j) + V_i \sum_{j \in i} B_{ij} V_j \sin(\theta_i - \theta_j), & i=n+1,\ldots,n+m \\ 0 = V_i \sum_{j \in i} B_{ij} E_j \sin(\theta_i - \delta_j) + V_i \sum_{j \in i} B_{ij} V_j \sin(\theta_i - \theta_j), & \text{the others} \end{cases} \quad (4)$$

where $E_i$ and $P_{Ei}$ are the $i$th generator's internal voltage and active power; $V_i$ and $\theta_i$ are the $i$th network nodal voltage amplitude and angle; $B_{ij}>0 (i \neq j)$ is the susceptance between bus $i$ and bus $j$, $B_{ii}=-\Sigma_{j \neq i} B_{ij}$; $P_{Li}$ is load's active power consumption for the $i$th bus.

To simplify the analysis, we made some assumptions: a) voltage magnitudes $E_i$ and $V_i$ are constant and equal to 1 p.u.; b) transmission lines are purely inductive; c) equilibrium angle differences are small enough.

Besides, we assume that all generators (their dynamics refer to (1), (2)) are similar [8]-[10]. That is, the turbine time constant $T_i$ are the same and their inertia $J_i$, damping coefficient $D_i$ and droop coefficient $K_i$ are in proportional, given as

$$\begin{cases} T_i = T \\ J_i : D_i : K_i = J_{sum} : D_{sum} : K_{sum} \end{cases} \begin{vmatrix} d = D_{sum}/J_{sum} \\ k = K_{sum}/J_{sum} \end{vmatrix}, i=1,2,\ldots,n \quad (5)$$

where $J_{sum} = \Sigma J_i$, $D_{sum} = \Sigma D_i$, $K_{sum} = \Sigma K_i$.

According to the above assumptions, linearizing (1), (4), and performing Laplace transform obtains the transfer function matrix between power disturbances and angles

$$\begin{bmatrix} \Delta u_G(s) \\ \Delta u_L(s) \end{bmatrix} = \left( \frac{1}{\omega_0} \begin{bmatrix} Js^2 + Ds + (Ts+I_N)^{-1} Ks & \end{bmatrix} - B \right) \begin{bmatrix} \Delta \delta_G(s) \\ \Delta \theta_L(s) \end{bmatrix} \quad (6)$$

where $[\Delta u_G(s)\ \Delta u_L(s)]^T$ is the vector of power disturbance change; $[\Delta \delta_G(s)\ \Delta \theta_L(s)]^T$ is the vector of angle change. Subscript $G$ indicates generators' internal buses and subscript $L$ indicates network buses. $J=\text{diag}(J_i)$, $D=\text{diag}(D_i)$, $T=\text{diag}(T_i)$, $K=\text{diag}(K_i)$ are diagonal matrices of inertia, damping coefficient, turbine time constant and droop coefficient respectively; $B$ is the network's susceptance matrix, $B_{GG} \in \mathbb{R}^{n \times n}$, $B_{GL} \in \mathbb{R}^{n \times (m+a)}$, $B_{LG} \in \mathbb{R}^{(m+a) \times n}$, $B_{LL} \in \mathbb{R}^{(m+a) \times (m+a)}$ are the susceptance submatrices of matrix $B$, given as

$$B = \begin{bmatrix} B_{GG} & B_{GL} \\ B_{LG} & B_{LL} \end{bmatrix} \quad (7)$$

Since the power step disturbance occurs at network buses, $\Delta u_G(s) = \mathbf{0}_{n \times 1}$, $\Delta u_L(s) = P/s$ in which $P=[P_i] \in \mathbb{R}^{(m+a) \times 1}$ is the vector of power disturbance's amplitude. Note that the elements of $\Delta \delta_G$ are state variables, the change of which is continuous; the elements of $\Delta \theta_L(s)$ are algebraic variables, the change of which can be abrupt. Based on this, it can be obtained from (6) with the constraints of the instantaneous active power balance in the network that

$$\Delta \theta_L(s) = \Delta \theta'_L(s) - B_{LL}^{-1} \Delta u_L(s) \quad (8)$$

where $\Delta \theta'_L(s)$ is the vector of network buses' angle change after the power step disturbance occurs.

Substituting (5), (8) into (6) obtains that

$$J^{\frac{1}{2}} \left( \frac{1}{\omega_0} \left( s+d+\frac{k}{Ts+1} \right) s I_N - J^{-\frac{1}{2}} B_r J^{-\frac{1}{2}} \right) J^{\frac{1}{2}} \Delta \delta_G(s) = -B_{GL} B_{LL}^{-1} \Delta u_L(s) \quad (9)$$

where $B_r = B_{GG} - B_{GL} B_{LL}^{-1} B_{LG}$ is bus-reduced susceptance matrix.

Since $B_r$ is a symmetric matrix and has a zero eigenvalue, $J^{-\frac{1}{2}} B_r J^{-\frac{1}{2}}$ is also a symmetric matrix, which has the characteristic that [9]-[12]

$$-U^T J^{-\frac{1}{2}} B_r J^{-\frac{1}{2}} U = \Lambda = \text{diag}\{\lambda_k\} \quad (10)$$

where $0=\lambda_1<\lambda_2<\ldots<\lambda_n$. $U=[U_1\ U_2\ \ldots\ U_n]$ and $U_1 = J_{sum}^{-\frac{1}{2}} J^{\frac{1}{2}} \mathbf{1}_{n \times 1}$. $\mathbf{1}_{n \times 1} \in \mathbb{R}^{n \times 1}$ is a vector, the elements of which are all one.

Substituting (10) into (9) obtains that

$$\Delta \delta_G(s) = -J^{-\frac{1}{2}} U \text{diag}\{\omega_0 H_k(s)\} U^T J^{-\frac{1}{2}} B_{GL} B_{LL}^{-1} \Delta u_L(s) \quad (11)$$

where $H_k(s) = \dfrac{Ts+1}{Ts^3 + (Td+1)s^2 + (d+k+\omega_0 T \lambda_k)s + \omega_0 \lambda_k}$.

Substituting (8) into (6) obtains that

$$\Delta \theta'_L(s) = -B_{LL}^{-1} B_{LG} \Delta \delta_G(s) \quad (12)$$

Equation (12) shows that $\Delta \theta'_L(s)$ is the vector of the superposition of generators' power angles $\Delta \delta_{Gi}$ ($i=1,\ldots,n$). Since $\Delta \delta_{Gi}$ has first-order derivative $\Delta \omega_{Gi}$, and second-order derivative $d\Delta \omega_{Gi}/dt$, it can be concluded that the first-order derivative of $\Delta \theta'_L(s)$ (i.e., the frequency change) and second-order derivative of $\Delta \theta'_L(s)$ (i.e., the derivative of the frequency change) also exist. Define the vector of frequency change of network buses as $\Delta \omega'_L(s)$. The expression of $\Delta \omega'_L(s)$ can be given as

$$\Delta \omega'_L(s) = s \Delta \theta'_L(s)/\omega_0 \quad (13)$$

Combing (11)-(13), we can obtain the transfer function matrix between $\Delta \omega'_L(s)$ and $\Delta u_L(s)$

$$\begin{aligned} \Delta \omega'_L(s) &= B_{LL}^{-1} B_{LG} J^{-\frac{1}{2}} U \text{diag}\{s H_k(s)\} U^T J^{-\frac{1}{2}} B_{GL} B_{LL}^{-1} \Delta u_L(s) \\ &= B_{LL}^{-1} B_{LG} J^{-\frac{1}{2}} \sum_{k=1}^{n} (U_k U_k^T s H_k(s)) J^{-\frac{1}{2}} B_{GL} B_{LL}^{-1} \Delta u_L(s) \quad (14) \\ &= \sum_{k=1}^{n} C_k \Delta u_L(s) s H_k(s) \end{aligned}$$

where $C_k = B_{LL}^{-1} B_{LG} J^{-\frac{1}{2}} U_k U_k^T J^{-\frac{1}{2}} B_{GL} B_{LL}^{-1} = [c_{k,ij}] \in \mathbb{R}^{(m+a) \times (m+a)}$.

Substituting $\Delta u_L(s) = P/s$ into (14) obtains that

$$\Delta\omega'_L(s) = \sum_{k=1}^{n} C_k P H_k(s) \quad (15)$$

According to (15), the initial RoCoF of network buses can be obtained by the initial value theorem

$$\left.\frac{d\Delta\omega'_L(t)}{dt}\right|_{t=0^+} = \lim_{s\to\infty} s^2 \Delta\omega'_L(s) = \left(\sum_{k=1}^{n} C_k\right) P = CP \quad (16)$$

where $C = [c_{ij}] \in \mathbb{R}^{(m+a)\times(m+a)}$ is a parameter matrix.

It can be concluded from (16) that the $i$th diagonal element of the parameter matrix $c_{ii}$ represents the proportional parameter of the initial RoCoF of the $i$th network bus under a power step disturbance occurring at the corresponding bus. Similar as (3), a metric is proposed to quantify the nodal frequency performance of the $i$th network bus.

**Definition**: a metric to evaluate nodal frequency performance of the $i$th network bus is defined as $1/c_{ii}$.

Note that the metric is related with the network structure and the generators' inertia. In the following, we will discuss the relation between the proposed metric and the nodal frequency performance of network buses.

Considering the power step disturbance occurs at the $i$th network bus only, it can be deduced from (15) that

$$\Delta\omega'_{Li}(s) = \sum_{k=1}^{n} c_{k,ii} H_k(s) P_i, \quad c_{ii} = \sum_{k=1}^{n} c_{k,ii} \quad (17)$$

Equation (17) shows that the dynamics of nodal frequency can be considered as the superposition of the $H_k(s)$ ($k=1,\ldots,n$). Moreover, it can be concluded that $H_k(s)$ ($k=1,\ldots,n$) are the frequency's dynamics of the $n$ decomposed single-machine system [9]-[10], which can be classified as two types:

a) when $k=1$, $\lambda_1=0$, $H_1(s)$ can be expressed as

$$H_1(s) = \frac{Ts+1}{Ts^3 + (Td+1)s^2 + (d+k+\omega_0 T\lambda_k)s} \quad (18)$$

b) when $k>1$, $H_k(s)$ can be expressed as

$$H_k(s) = \frac{Ts+1}{Ts^3 + (Td+1)s^2 + (d+k+\omega_0 T\lambda_k)s + \omega_0 \lambda_k} \quad (19)$$

Note that $H_1(s)$ reflects the dynamics of the system frequency (COI frequency), $H_k(s)$ reflects the nodal frequency deviation from the system frequency [8].

Besides, $c_{ii}$ is expressed as

$$c_{ii} = c_{1,ii} + \sum_{k=2}^{n} c_{k,ii} \quad (20)$$

According to (10) and (14), $c_{1,ii}$ can be deduced as

$$c_{1,ii} = 1/J_{sum} \quad (21)$$

Equation (21) shows that $c_{1,ii}$ is a constant value independent from the network buses. Since $c_{k,ii}$ is the proportional parameter of $H_k(s)$ for the frequency dynamics of the $i$th network bus in (17), it can be considered from (20) that $c_{ii}$ can denote the impact of the frequency deviation components $H_k(s)$ ($k>1$) on the $i$th network nodal frequency. That is, if $c_{ii}$ is larger, the frequency deviation components have a larger influence on the $i$th network nodal frequency. This indicates that the proposed metric can evaluate the nodal frequency performance of network buses, i.e., if the proposed metric ($1/c_{ii}$) is larger, the nodal frequency performance of the $i$th network bus is better.

As an example, a two-generator power system is conducted to introduce the proposed metric.

**B. Example**

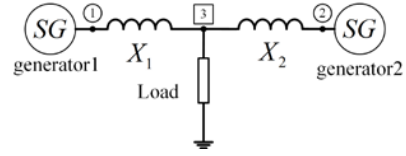

Fig. 1 Two-generator power system with one network bus

Fig. 1 show a two-generator power system, where bus 1, 2 are the generators' internal buses, bus 3 is a network bus with a load, $X_1$, $X_2$ are reactance. Inertia of the two generators are $J_1 = 20$p.u. and $J_2 = 10$p.u. respectively. The susceptance matrix of this network is given as

$$B = \begin{bmatrix} B_{GG} & B_{GL} \\ B_{LG} & B_{LL} \end{bmatrix} = \begin{bmatrix} -\frac{1}{X_1} & & \frac{1}{X_1} \\ & -\frac{1}{X_2} & \frac{1}{X_2} \\ \frac{1}{X_1} & \frac{1}{X_2} & -\frac{1}{X_1} - \frac{1}{X_2} \end{bmatrix} \quad (22)$$

Substituting (22) into (14) and (16) obtains proposed metric $1/c_{33}$ and its components $1/c_{1,33}$, $1/c_{2,33}$

$$\begin{cases} 1/c_{33} = \dfrac{1}{\dfrac{1}{J_1}\left(\dfrac{X_2}{X_1+X_2}\right)^2 + \dfrac{1}{J_2}\left(\dfrac{X_1}{X_1+X_2}\right)^2} \\ 1/c_{1,33} = J_1 + J_2 \\ 1/c_{2,33} = (J_1 + J_2)\dfrac{(X_1+X_2)^2}{\left(\sqrt{\dfrac{J_2}{J_1}}X_2 - \sqrt{\dfrac{J_1}{J_2}}X_1\right)^2} \end{cases} \quad (23)$$

According to (23), the propose metric is related with generators' inertia and network structure.

Fig. 2 illustrates the loci of $1/c_{33}$, $1/c_{1,33}$ and $1/c_{2,33}$ with varying position of bus 3, i.e., the reactance ratio $X_1/(X_1+X_2)$.

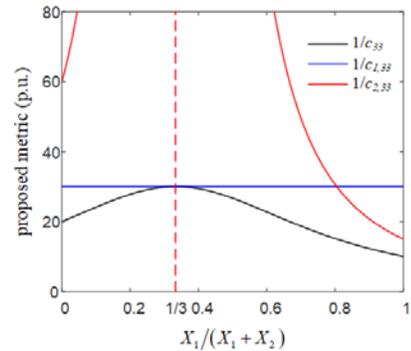

Fig. 2 Propoesed metric's loci

Fig. 2 shows that if the network bus is closer to a generator's internal bus, the metric of this network bus will tend to be the generator's inertia, e.g., if $X_1/(X_1+X_2)=0$, then $1/c_{33} = J_1 = 20$p.u.. Besides, Fig. 2 shows that the proposed metric gets maximum when bus 3 locates at a certain position between bus 1 and bus 2, e.g., $X_1/(X_1+X_2) = 1/3$ for this case.

## IV. CASE STUDY

In this section, a four-generator power system, as shown in Fig. 3, is provided to verify the proposed metrics. The parameters of generators and network are given in TABLE 1. Bus1-4 are generators' internal buses, bus 5-8 are network buses.

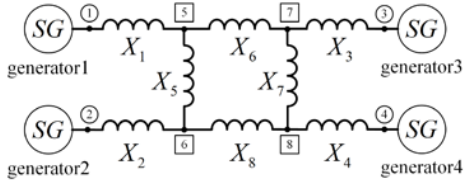

Fig. 3 Four-generator power system with four network buses

TABLE 1      PAPAMETER OF THE FOUR-GENERATOR POWER SYSTEM

| parameter | value |
|---|---|
| $J_1, J_2, J_3, J_4$ | 12, 8, 8, 4 (p.u.) |
| $D_1, D_2, D_3, D_4$ | 1.5, 1, 1, 0.5 (p.u.) |
| $K_1, K_2, K_3, K_4$ | 18, 12, 12, 6 (p.u.) |
| $T_1, T_2, T_3, T_4$ | 7, 7, 7, 7 (s) |
| $X_1, X_2, X_3, X_4$ | 0.1, 0.1, 0.1, 0.1 (p.u.) |
| $X_5, X_6, X_7, X_8$ | 0.2, 0.4, 0.4, 0.6 (p.u.) |

TABLE 2 shows the proposed metrics of the network buses. It can be seen from TABLE 2 that the metrics of network buses have the order from large to small are bus 5, 6, 7, 8. This indicates the oscillating level of frequency deviations from small to large have same order. Besides, TABLE 2 shows that the $1/c_{1,ii}$ ($i=5,6,7,8$) for each network buses are all the same, which indicate that the network buses have the same system frequency response.

TABLE 2      PROPOSED METRIC AND THEIR COMPONENTS

| bus number | $1/c_{ii}$ | $1/c_{1,ii}$ | $1/c_{2,ii}$ | $1/c_{3,ii}$ | $1/c_{4,ii}$ |
|---|---|---|---|---|---|
| 5 | 24.0 | 32.0 | 148.8 | 276.4 | 2.2e6 |
| 6 | 16.3 | 32.0 | 160.3 | 45.1 | 653.2 |
| 7 | 14.7 | 32.0 | 30.7 | 59.3 | 378.9 |
| 8 | 7.2 | 32.0 | 56.3 | 91.2 | 12.4 |
| 8* | 13.8 | 32.0 | 36.3 | 74.3 | 4.7e3 |
| 8** | 14.0 | 32.0 | 45.1 | 77.8 | 190.8 |

\* exchange generator 2 with generator 4

\*\* changing $X_4$ from 0.1p.u. to 0.3p.u.

At $t = 5$s, a power step disturbance $\Delta u = -0.1$p.u. occurs at bus 5-8, respectively. The frequency responses of the corresponding bus are shown in Fig. 4. Fig. 4 shows that the frequency responses for the four cases are similar, which indicates that the system frequency component are the same. Besides, it can be seen from Fig. 4 that the order of frequency deviation oscillating level consistent with the theoretical analysis results.

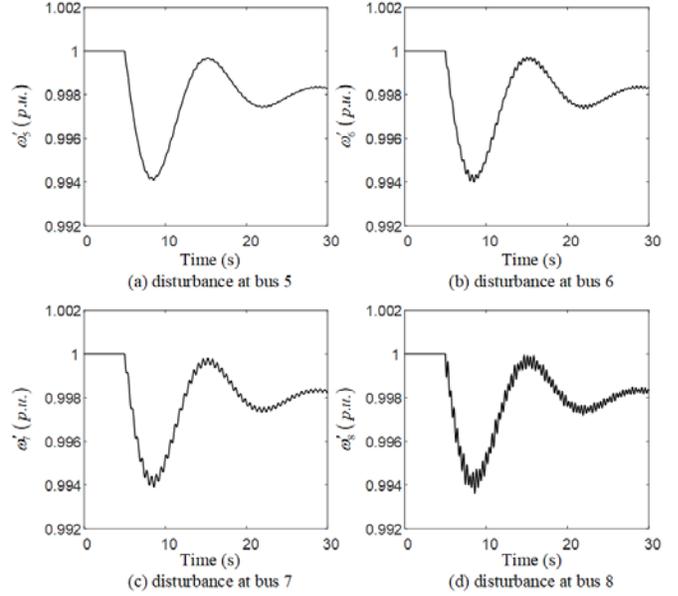

Fig. 4 Frequency responses of bus 5-8

Besides, another two cases are provided: case a) exchange generator 2 with generator 4; case b) increase reactance $X_4$ from 0.1 p.u. to 0.3 p.u.. The corresponding metrics of bus 8 refer to Table 2. For case a, it can be seen from Table 2 that the metric of bus 8 becomes better after the exchange. This indicates that proper inertia distribution can improve nodal frequency performance. For case b, TABLE 2 shows that the increase of reactance $X_4$ improves the metric of bus 8. The reason is that the inertia of generator 4 is the smallest in this four-generator power system. Besides, according to the discussion of the two-generator power system in Section III, when the network bus gets closer to a generator, the metric of this network bus will get closer to the generator's inertia. Therefore, if the network bus is further from generator 4, the metric of this network bus will be larger. The frequency responses of bus 8 under a power step disturbance $\Delta u = -0.1$p.u. are shown in Fig. 5 for these two cases. Fig. 5 (a), (b) both show that the frequency response has been improved in comparison with Fig. 4 (d). The simulation results verify the theoretical analysis.

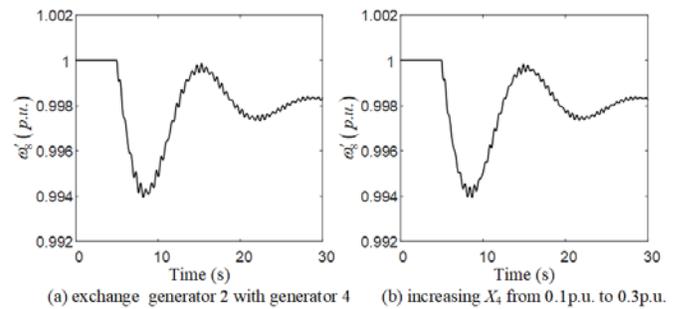

Fig. 5 Frequency responses of bus 8

As a result, these cases prove that proposed metric can evaluate the nodal frequency performance. Besides, it should be mentioned that strengthen network doesn't always means better proposed metric. The nodal frequency performance is also related with the system's inertia distribution. The relation of the proposed metrics with system's inertia distribution and network structure need further research.

V. CONCLUSION

Inspired by the relation between the inertia and initial RoCoF in generators, this paper presents how the initial RoCoF of the nodal frequencies are related to the inertia constants of multiple generators in a power network, which leads to a performance metric to access nodal frequency performance. The proposed metric can represent the impact of frequency deviation components on frequency dynamics for network buses. A larger value of this metric indicates that the frequency performance of the corresponding network bus is better. Simulation results verify the validity of the proposed metric. Our future works will focus on the relationship between generator distribution, network structure and the metric.